\documentclass[twocolumn,secnumarabic,floatfix,amssymb,amsmath,nofootinbib,tightenlines,showpacs,superscriptaddress,aps,prl]{revtex4-1}
\usepackage{amsfonts}
\usepackage[colorlinks=true,linkcolor=blue,pagecolor=blue,filecolor=blue,menucolor=yellow,urlcolor=blue,citecolor=blue,anchorcolor=blue]{hyperref}%
\usepackage{graphicx}
\usepackage{dcolumn}
\usepackage{bm}
\usepackage{times}
\usepackage{sidecap}
\usepackage{float}
\usepackage{soul}
\newcommand{\etal}{\emph{et al.}}

\begin{document}

\title{Spin-controlled superconductivity and tunable
triplet correlations in graphene nanostructures} 

\author{Klaus Halterman }
\email{klaus.halterman@navy.mil}
\affiliation{Michelson Lab, Physics
Division, Naval Air Warfare Center, China Lake, California 93555}
\author{Oriol T. Valls}
\email{otvalls@umn.edu}
\altaffiliation{Also at Minnesota
Supercomputer Institute, University of Minnesota, Minneapolis,
Minnesota 55455} \affiliation{School of Physics and Astronomy,
University of Minnesota, Minneapolis, Minnesota 55455}
\author{Mohammad Alidoust}
\email{phymalidoust@gmail.com}
\altaffiliation{Also at Department of
Physics, Faculty of Sciences, University of Isfahan, Hezar Jerib
Ave, Isfahan 81746-73441,Iran} \affiliation{Department of Physics,
Norwegian University of Science and Technology, N-7491 Trondheim,
Norway}

\date{\today}

\begin{abstract}
We study 
graphene
ferromagnet/superconductor/ferromagnet (F/S/F) nanostructures 
via a microscopic self-consistent Dirac Bogoliubov-de Gennes
formalism. We show that as a result of proximity effects, 
experimentally accessible spin switching phenomena can
occur as one tunes the Fermi level $\mu_F$ of the F regions or
varies the angle $\theta$ between exchange field orientations. 
Superconductivity can then be switched on and
off by varying either $\theta$ or $\mu_F$ (a
spin-controlled superconducting graphene switch). 
The
induced equal-spin triplet correlations in  S  can be 
controlled by tuning $\mu_F$, effectively making a graphene based
two-dimensional spin-triplet valve.
\end{abstract}
\maketitle

\noindent \emph{Introduction.---}The continual development of
graphene has sparked vast research efforts
linked to emerging nanotechnologies that span numerous scientific
disciplines. With its 2D hexagonal lattice structure 
and linear energy dispersion at low energies, graphene
possesses many desirable properties \cite{novo,son,kat}, such as 
extremely high electrical conductivity, tensile strength, and
thermal conductivity. Graphene has  played a prominent part 
in  recent significant advances involving 
transistors, solar cells, and tunable THz electromagnetic radiation
detection \cite{vica}. Although graphene is intrinsically a gapless
semiconductor, it can  become superconducting (SC) as well
as acquire ferromagnetic (FM) properties through doping or
defects
\cite{son,Berger}. 
Thus, with the existing capability to create hybrid structures
involving  FM and
 SC graphene, researchers are now
seeking breakthroughs involving graphene-based low temperature
spintronic devices. The effectiveness of a graphene spin valve or
switch involving both ferromagnet (F) and
superconductor (S) elements in contact is based on 
proximity effects, which dictate the behavior of both the singlet
and triplet pairing correlations in each region
\cite{linder1,hsu,bum,Tworzydo,Liang,Zhang}.
In graphene,  as opposed to most 
conventional materials, the Fermi level can be tuned via a gate
potential. This leads to many practical 
SC device applications, e.g, the supercurrent in a nonmagnetic
Josephson junction can be reversed by tuning the gate voltage
\cite{Liang}.

In this Letter, we demonstrate that pairing correlations in F/S/F
graphene structures can be experimentally modulated in a
controllable way by changing either the relative magnetization
direction or by manipulating the Fermi level $\mu_F$ of 
the F layers, leading to new possibilities for graphene-based
devices.
\begin{figure}
\includegraphics[width=7.3cm,height=3.5cm] {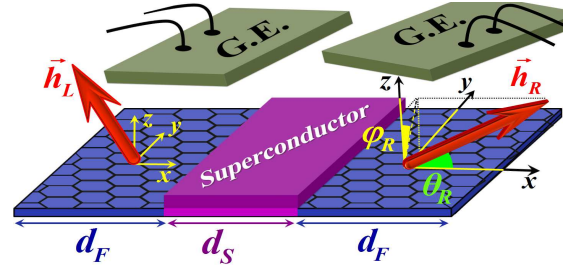} 
\caption{(Color online) F/S/F geometry
investigated. The magnetic structure is described by
exchange fields $\vec{\textit{\textbf{h}}}_L$ and 
$\vec{\textit{\textbf{h}}}_R$ in the left and right magnets
respectively. Their orientation is given 
by angles $\theta_{L,R}$ and $\varphi_{L,R}$:
$\vec{\textit{\textbf{h}}}_{L,R}$
=$h_{L,R}$($\cos\theta_{L,R}$,$\sin\theta_{L,R}$$\sin\varphi_{L,R}$,$
\sin\theta_{L,R}$ $\cos\varphi_{L,R}$).
The Fermi level in the F regions can be tuned via the gate
electrodes (G.E.). The system is infinite in the $x$-$y$ plane.
$x$ is normal to the interfaces.} 
\label{fig}
\end{figure}
We show that by tuning $\mu_F$ (which can be done e.g. by applying
an external electric field), a new type of spin switching can occur
where the system transitions between a resistive normal state and a 
 SC one. This spin switching  is closely
linked to the magnitude of the exchange interaction in the F layers:
in  FM graphene the exchange field, $h$,
shifts the Dirac points for spin-up and spin-down quasiparticles by
an amount $-h$ and $+h$ respectively. This shift can create an
effective barrier
for the electrons and holes, and its existence 
relates to spin dependent Klein
tunneling \cite{son,kat}. We find also that 
similar transitions between the resistive  and
SC states can be realized by variation
the relative  angle between the
field orientations of the two F regions.   
In the graphene F/S/F system studied, equal-spin triplets can
arise and be utilized as an effective triplet spin transistor: for
certain exchange field 
configurations, the amount of equal-spin pairs
in the superconductor can be controlled by the Fermi levels in the
ferromagnets.

\noindent \emph{Method.---}To study these phenomena, we have
developed a microscopic self-consistent Dirac Bogoliubov-de 
Gennes (DBdG) approach to characterize  
graphene-based systems involving $s$-wave superconductors and
ferromagnets with general exchange field 
orientations and tunable
Fermi levels.
We also computed the critical temperature  of a F/S/F nanostructure
from the linearized\cite{hvagraph} self consistency equation. The
confinement of the massless Dirac fermions along 
the $x$ direction in Fig.~\ref{fig} requires boundary conditions
compatible with the DBdG equations. As in the famously difficult
infinite potential well problem in relativistic quantum mechanics
\cite{berry}, the vanishing of the wavefunction at the outer
boundaries of a quasi-1D ``box" becomes problematic and 
alternate approaches are needed to avoid 
manifestations of the Klein paradox \cite{kat}. Several approaches
are possible. The one we take here (analogous to methods 
used in the study of the ``bag model") incorporates 
in the Hamiltonian terms corresponding to fictitious additional
layers in the outer regions beyond the studied system \cite{berry,uslater}.

The inclusion of 
magnetism into the DBdG Hamiltonian yields the following
16$\times$16 Hamiltonian where $\pm$-signs refer to the graphene
valleys, $K (+)$ and $K' (-)$ \cite{hsu,Tworzydo,Liang,Zhang}:
\begin{eqnarray}\label{dbdg}
&\left(
  \begin{array}{cc}
    \mathcal{H}_M-\mu\hat{I} & \Delta(\vec{r}) \hat{I}\\
    \Delta^{\ast}(\vec{r}) \hat{I} & -(\mathcal{T}\mathcal{H}_M\mathcal{T}^{-1}-\mu\hat{I}) \\
  \end{array}
\right) \Psi_n =\epsilon_n \Psi_n,\\&\nonumber {\cal H}_M={\cal
H}\otimes\sigma_0-\sigma_0\otimes
\vec{\textit{\textbf{h}}}_{L,R}\cdot\vec{\mathbf{\sigma}},\;\;\;\;\;\;{\cal
H}=
\begin{pmatrix}
{\cal H}^+&0 \\
0&{\cal H}^-
\end{pmatrix},
\end{eqnarray}
where $\hat{I}$ is the 8$\times$8 identity matrix, $\mu$  the
chemical potential, and $\mathcal{T}$  the time reversal operator.
The 1$\times$16 wavefunction is $\Psi_n \equiv
(\Psi^\uparrow_{u,n},\Psi^\downarrow_{u,n},\Psi^\downarrow_{v,n},\Psi^\uparrow_{v,n})$,
with $\Psi^\sigma_{u,n} \equiv
(u^\sigma_{n,A,K}$,$u^\sigma_{n,B,K}$,$u^\sigma_{n,A,K'}$,$u^\sigma_{n,B,K'})$,
and $\Psi^\sigma_{v,n} \equiv
(v^\sigma_{n,A,K'}$,$v^\sigma_{n,B,K'}$,$v^\sigma_{n,A,K}$,
$v^\sigma_{n,B,K})$, with $\epsilon_n$ being the Dirac fermions
eigenenergies. The labels $A$ and $B$ denote the two sublattices
that arise from the honeycomb lattice structure. The chemical
potential takes the value $\mu_F$ in the two magnets and $\mu_S$ in
the superconductor. The time-reversal operator in our chosen basis
is given by $\mathcal{T} = [\sigma_z \otimes \sigma_x \otimes (-i
\sigma_y)] \mathcal{C}$, where $\mathcal{C}$ is the  complex
conjugation operator. ${\cal H}^\pm$ is  written succinctly as
${\cal H}^\pm = v_F (\sigma_x p_x \pm \sigma_y p_y)+v_F^2 M(x)
\sigma_z$. Here $\sigma_i$ are the 2$\times$2 Pauli matrices acting
in sublattice space, $\sigma_0$ the 4$\times$4 identity matrix, and
$v_F$ is the 
Fermi velocity in graphene.
We have introduced, to avoid the previously mentiond
Klein paradox \cite{kat,berry} problems, 
a  mass term $M(x)$ \cite{Tworzydo} which vanishes in the F/S/F
region studied and is effectively infinite in two computationally
added outer regions adjacent to the F portions \cite{uslater}.
The exchange fields, 
$\vec{\textit{\textbf{h}}}_{L,R}$,  in the left ($L$) and right
($R$) F electrodes  can be of different magnitude and have 
different orientation angles (see Fig.\ref{fig}). 
We will present results here only for 
the case where $\phi_{L,R}=\pi/2$, 
$\theta_R=0$ and equal magnitudes, $h_L=h_R\equiv h$. 

The coupling of electrons in a given valley with the  hole
excitations in the other one is accomplished through the $s$-wave pair
potential $\Delta(x)$ \cite{Tworzydo}, and is determined
self-consistently by
$\Delta(x)$=$  {g}/{2}{\sum_{n,\lambda,\beta}}
(u^\uparrow_{n,\lambda,\beta} v^{\downarrow
*}_{n,\lambda,\bar{\beta}}
 + u^\downarrow_{n,\lambda,\beta} v^{\uparrow *}_{n,\lambda,\bar{\beta}}
) \tanh({\epsilon_n}/{2T})$,
where $\lambda$ represents the sublattice index ($A$ or $B$),
$\beta$ the valley index ($K$ or $K'$), and $T$ the system
temperature. The coupling parameter, $g$, is a constant finite only
in the superconductor region. The sum is restricted to those quantum
states with positive energies below a ``Debye" energy cutoff,
$\omega_D$.
With the quantization axis aligned along the $z$ direction, the
triplet amplitudes, $f_{0,z}$ and $f_{1,z}$, can be written as 
$f_{0,z} =  1/2\sum_{n}(f_n^{\uparrow\downarrow}-f_n^{\downarrow\uparrow}) \zeta_n(t)$,
and $f_{1,z}  =1/2\sum_{n} (f_n^{\uparrow\uparrow}+f_n^{\downarrow\downarrow})\zeta_n(t)$,
where $\zeta_n(t) = \cos(\epsilon_n t)-i\sin(\epsilon_n
t)\tanh(\epsilon_n/2 T)$, and we define $f_{n}^{\sigma\sigma'}
\equiv   \sum_{\lambda} [ u^\sigma_{n,\lambda,K} v^{\sigma'\ast
}_{n,\lambda,K'} + u^\sigma_{n,\lambda,K'} v^{\sigma'\ast
}_{n,\lambda,K}] $ \cite{hbv}.
For structures such as ours where the direction of the exchange
fields varies with position
it is more
insightful to align the quantization axis with the local
field vector. This  helps distinguish the long-range 
nature of the equal-spin triplet correlations ($f_1$) from the
damped oscillatory behavior of the opposite-spin triplets ($f_0$). 
This is achieved by  performing the appropriate spin rotations \cite{uslater}.  
For the orientations considered
here, the rotated amplitudes are
$f_0^{L,R} =  {1}/{2}\sum_{n} 
\lbrace\cos\theta_{L,R}(f_n^{\uparrow \uparrow} +
f_n^{\downarrow \downarrow})
+i (f_n^{\uparrow \uparrow} - f_n^{\downarrow \downarrow}) 
\rbrace\zeta_n(t) $ and 
$f_1^{L,R}  ={1}/{2}\sum_{n} 
\lbrace\-\sin\theta_{L,R}(f_n^{\uparrow \uparrow} +
f_n^{\downarrow \downarrow}) +i (f_n^{\uparrow
\uparrow} -
f_n^{\downarrow \downarrow}) 
\rbrace\zeta_n(t)$, where one sets $\theta \equiv \theta_L$ on the left
side and $\theta_R=0$ for the right side.
The singlet pair amplitude is of course invariant under these rotations.
When the exchange fields 
lie along one of the coordinate axes, 
the {\it singlet} pair amplitude depends only
on whether the relative exchange field orientations are parallel, 
antiparallel, or perpendicular to each other \cite{uslater}. 
The  self-consistent methods used are extensions  
of those previously published\cite{hvagraph}  but now 
the matrix dimensions involved are doubled by
inclusion of the
spin degree of freedom. 

\noindent \emph{Results.---}
As stated above, we assume that the exchange fields
have the same magnitude, $h$ (see
Fig.~\ref{fig}), and lie in
the plane of the graphene 
($\varphi_{L,R} = 90^\circ$ and $\theta_R=0^\circ$). Thus the
exchange field orientation is  described by the angle 
$\theta_L \equiv \theta$ 
which can be manipulated  by the transfer of spin angular
momentum from injected electrons or by an external magnetic field.
All spatial quantities are scaled by the Fermi 
wavevector, $k_{FS} \equiv \mu_S/v_{FS}$. We take $d_F$ and $d_S$ to
be the same and (in scaled units) to equal the dimensionless
SC coherence length: $k_F \xi_0 = 100$,
where $\xi_0 = v_{FS}/\Delta_0$. Thus, in the figures the S region
lies in the range 200-300 when spatially dependent quantities are
shown. In the results that follow, $h$ is normalized by $\mu_S$ and
we consider the ratio $\widetilde{\mu}_F\equiv \mu_F/\mu_S$ when
describing the relative Fermi levels.

\begin{figure}
\includegraphics[width=8.65cm,height=3.6cm] {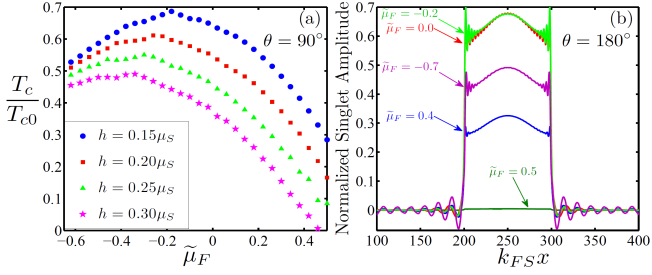}
\caption{(Color online) (a): Critical temperature $T_c$ (normalized
by its bulk value $T_{c0}$) vs. the magnet doping parameter
$\widetilde{\mu}_F$ for various values of $h/\mu_S$ with $\theta =
90^\circ$ (orthogonal exchange fields). In (b) the singlet pair
amplitude at $T/T_{c0} = 0.29$ is shown as a function of position, 
for an exchange field $h/\mu_S=0.2$, and  with $\theta = 180^\circ$
(antiparallel exchange fields). Results for several 
$\widetilde{\mu}_F$ are
shown.} \label{tcmu} 
\end{figure}

Figure \ref{tcmu} displays the gate voltage switching effect. The
critical temperature, $T_c$, (computed by generalizing
the methods of Ref.~\cite{hvagraph} and normalized to its value in a pure S
sample) of the  system is plotted in panel (a) as a function of
Fermi level shift for various values of the exchange field. 
The results in this panel correspond
to a perpendicular ($\theta=90^\circ$) configuration.
As can be seen, $T_c$ is nonmonotonic: it is
largest when $\widetilde{\mu}_F$ is near $-h/\mu_S$. At larger values of
$\widetilde{\mu}_F$, $T_c$ decreases sharply:
for $T \gtrsim T_c$, increases in
$\widetilde{\mu}_F$ at constant $\theta$ and $h/\mu_S$
switch the system from 
 SC to normal. When
$|{\mu}_F| \gg h$, 
superconductivity 
becomes less dependent on the exchange field as evidenced by
the coalescing of the curves for 
larger negative values of $\widetilde{\mu}_F$: 
the proximity effects
diminish due to the extreme mismatch in Fermi levels, 
resulting in greater isolation of the three regions. $T_c$, in turn
is weakly dependent on both $h$ and $\mu_F$. This behavior is also
found for large positive $\widetilde{\mu}_F$ (not shown). For
moderate $\widetilde{\mu}_F$, the self-consistent proximity effects
become even more important as the pair-breaking ferromagnet regions
strongly reduce the SC correlations,
resulting in the observed decline in $T_c$ towards zero. This
nontrivial behavior is further influenced by the shifting of the
Dirac points by the exchange field, causing a corresponding shift in
the peaks of each of the four curves in Fig.~\ref{tcmu}(a). These
features are absent in standard metals, which lack tunable Fermi
surfaces.
The degree to which superconductivity can be tuned
via $\widetilde{\mu}_F$ depends on 
the magnetic configuration of the system.
We show
in panel (b),
the antiparallel field configuration,  $\theta=180^\circ$. There, the 
spatial
profile of the self-consistent singlet pair amplitude (normalized to
its value in a pure S sample) is shown at several values of
$\widetilde{\mu}_F$, at $T/T_{c0} = 0.29$. Just as for  the $T_c$ 
results in (a), we find nonmonotonic behavior in
$\widetilde{\mu}_F$: for $\widetilde{\mu}_F < -h/\mu_S$, the 
singlet correlations are enhanced in the S region as the
Fermi level shift increases.
However if $\widetilde{\mu}_F$ is  increased beyond about
$-h/\mu_S$, SC correlations rapidly
decrease and
vanish 
as  $\widetilde{\mu}_F$ reaches  0.5. Thus, by tuning the relative
Fermi level of the F regions (e.g., by an electric field), the
system will switch from a  SC state to
normal one (or vice versa) depending on the
field
configuration. 

\begin{figure}
\includegraphics[width=8.65cm,height=3.6cm] {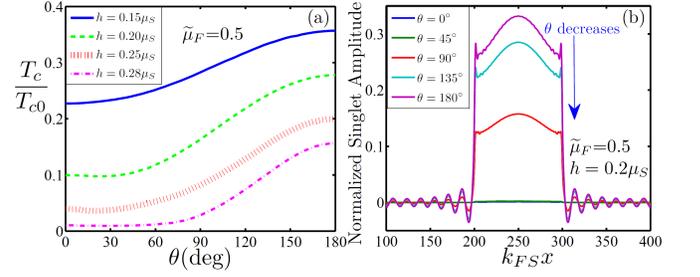}
\caption{(Color online) (a): Critical temperature versus the
relative field angle $\theta$ at several
exchange fields,
$h/\mu_S$.
The doping
parameter $\widetilde{\mu}_F$ is fixed at $\widetilde{\mu}_F =0.5$.
In (b) we show the singlet pair amplitude normalized to its
bulk value  for different field
orientations, $\theta$ (see legend). $T$ is set at
$T/T_{C0} = 0.12$, and $\widetilde{\mu}_F  = 0.5$. }
\label{tctheta}
\end{figure}

In Fig. \ref{tctheta} we exhibit the switching of superconductivity
with  orientation angle $\theta$. 
Panel (a) displays
the normalized $T_c$ vs. $\theta$ at
$\widetilde{\mu}_F=0.5$, for several values 
of $h$. 
For all four exchange fields shown, $T_c$ increases monotonically 
as $\theta$ goes from the parallel ($\theta = 0^\circ$) to the 
antiparallel ($\theta = 180^\circ$) configuration. 
Increasing the exchange
field results in greater pair-breaking and thus an overall reduction
in $T_c$.
The curves  are not related by a simple shift or factor,
reflecting the nontrivial self-consistent nature of the solutions. 
$T_c$ can be increased if one of the F layers is hole-doped while the other is electron-doped,
but its sensitivity to $\theta$ is much less. 
The  greatest difference between the  $T_c$
values of parallel and antiparallel states occurs here for intermediate
$h\approx0.2 \mu_S$.
We also see from Fig.~\ref{tctheta}(a) that
when $\theta=180^\circ$ and $h=0.2 \mu_S$,   $T_c/T_{c0}\approx 0.28$,
which is consistent with Fig.~\ref{tcmu}(b), where at $T/T_{c0} = 0.29$,
and $\widetilde{\mu}_F=0.5$,
the system is slightly above $T_c$. 

If $T$ is just above the minimum in the $T_c$ vs. $\theta$
curves, superconductivity can be switched on by increasing $\theta$.
The angular dependence of 
$T_c$ found here is 
consistent with recent experimental and theoretical observed trends
in 3D metallic  F/S/F trilayers \cite{Zhu}. Panel (b) demonstrates
the spatial dependence of the zero temperature self-consistent
singlet pair amplitude for a representative set of
field orientations.
The behavior of this amplitude in the F regions reflects 
the characteristic damped oscillations arising from the spin
splitting effects of the magnetism. As $\theta$ decreases,
superconductivity declines dramatically showing again that by
controlling the  field orientation, it can be
switched on or off. Thus, Figs.~\ref{tcmu} and \ref{tctheta}
describe a graphene-based F/S/F nanostructure with normal to
 SC  switching induced by variation of
either relative  field orientation or of the
Fermi level of the F regions.
If we consider opposite Fermi level shifts
in the FM regions i.e. $\mu_{FL}$=-$\mu_{FR}$, we find that
the observed switching phenomena disappear \cite{uslater}.

\begin{figure}
\includegraphics[width=8.65cm,height=3.4cm] {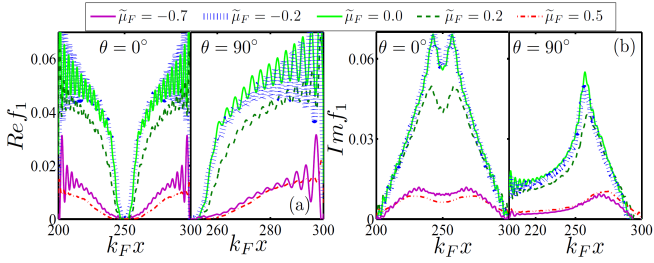}
\caption{(Color online)  Triplet spin valve effect. In panels (a)
and (b), respectively, we plot the real and imaginary parts of the
equal-spin triplet correlations, $f_1$, in the S region. The curves
in each of the panels represents a different doping level
$\widetilde{\mu}_F$ in the F regions: by tuning $\widetilde{\mu}_F$,
the degree of equal-spin correlations in the superconductor can be
controlled. The two relative exchange field orientations studied 
correspond to when the internal exchange fields are parallel ($\theta = 
0^\circ$) or perpendicular ($\theta = 90^\circ$) } \label{singlet}
\end{figure}

We now proceed to discuss the  induced
triplet pairs, which
generate appreciable spin-valve effects in these 
junctions. Since the triplet amplitudes are odd in time,
we must consider
finite relative time differences, $t$, in the triplet correlation functions.
We scale $t$ by the ``Debye" energy, $\omega_D$,  and choose
$\omega_Dt=4.0$ as a 
representative value to discuss the
behavior of the triplet amplitudes. 
We focus on moderately magnetic materials and set $h/\mu_S=0.2$.
The most interesting triplet amplitudes
are those with nonzero spin projection on the quantization axis.
Figure \ref{singlet} illustrates the  spatial
dependence of these spin-triplet
correlations in S. In panels (a)  and
(b) we plot respectively the
real and imaginary parts of the equal-spin
triplet amplitude $f_1$ 
in the S region for several values of
 $\widetilde{\mu}_F$.  Both parallel,
$\theta=0^\circ$, and perpendicular, $\theta=90^\circ$, relative
exchange field orientations are included. 
These plots show that the 
equal-spin triplets more readily populate S when both magnets are in
the parallel state ($\theta=0^\circ$). For a given $\theta$, the
scale of the triplet correlations in S is 
then governed by the
tunable Fermi shift in the F regions.
We have found that on the
other hand, in the F regions, the triplet amplitudes with zero  spin
projection (on the local quantization axis), $f_0$, are  spatially
characterized by the usual damped oscillatory behavior, similar to that of 
the ordinary singlet amplitudes.



\begin{figure}[t]
\includegraphics[width=8.65cm,height=3.4cm] {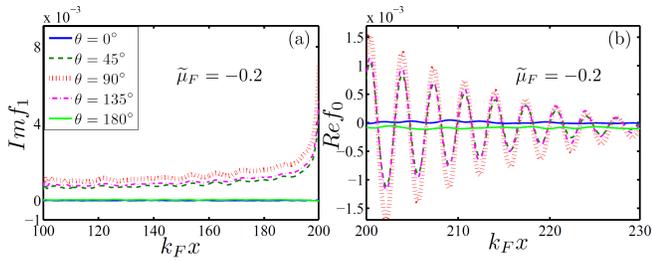}
\caption{(Color online) The imaginary part of $f_1$ in F (panel (a))
and the real part of $f_0$ in S (panel (b)) are plotted for multiple
values of $\theta$ and  $\widetilde{\mu}_F=-0.2$. (The corresponding
imaginary and real parts vanish). The long range nature of $f_1$ in
F is seen in panel (a) while the obvious contrast between the long-
and short-ranged nature of the equal- and opposite-spin triplets
($f_1$ and $f_0$) in S is seen by comparing panel (b)  with
Fig.~\ref{singlet}.
} \label{triplet}
\end{figure}
The contrast between the spatial behavior of $f_0$ and $f_1$ in the
F and S regions is best seen by comparing Figs.~\ref{singlet} and
\ref{triplet}.
In Fig.~\ref{triplet}, panels (a) and (b) display respectively the
imaginary part of $f_1$ in F and the real part of $f_0$ in S, for
several values of $\theta$ at $\widetilde{\mu}_S=-0.2$. Panel (a)
shows that $f_1$ is
long ranged in F and, when the relative exchange fields 
are noncollinear, it pervades the entire magnet
region. One also sees, in panel (b), that the opposite-spin triplets
are small and short ranged in the S region, also vanishing at
$\theta=0^\circ$, and $180^\circ$. In contrast, panels (a) and (b)
in Fig.~\ref{singlet} illustrate non-vanishing equal-spin
correlations in S at the same value of $\widetilde{\mu}_S$. Thus 
the graphene-based F/S/F nanostructure can be utilized
as a triplet spin valve. To achieve this 
equal-spin triplet
spin valve effect, at least one of the ferromagnets
should have a component of its magnetization 
out-of-plane. More important, the triplet spin switching aspect of
the system can be experimentally achieved by modulating either the
relative
field orientation or the Fermi level \cite{Zhu}. 
This tuning is unavailable in ordinary materials. 

\noindent \emph{Conclusions.---} We have shown that F/S/F graphene
nanostructures can exhibit very rich and experimentally accessible
spin switching phenomena: For particular values of the relative
field orientation of the two F regions, the induced
equal-spin triplet correlations in the S region can be
experimentally modulated in a controllable fashion by manipulating
the Fermi level. This is in turn suggestive of a carbon-based
spin-triplet transistor. 
Variations in relative  field orientations or
Fermi
levels of the F regions allow the superconductivity to be 
switched on and off, thus producing  a 
spin-controlled  SC graphene switch. 
The results presented here
are particular
to the Dirac-like band structure in graphene,
where, based on the magnetic configuration,
the Fermi level can be shifted
in a controllable fashion (by doping or electric fields).
They are also dependent upon the intrinsic 2D geometry,
where the confining boundaries result in
``relativistic" quantum interference effects
not present in ordinary 3D metals.
With recent experimental advances, 
including gate-tunable SC graphene
hybrids \cite{kessler}, this work should stimulate future
experiments involving graphene-based spin-switch devices. One
possibility could involve magnetoresistance measurements for a
sample configuration similar to that in Fig.~\ref{fig}, where the
gate electrodes control the local Fermi level. The predicted
spin-switch signatures should also be revealed in transport
experiments via  SC electrodes
\cite{heer}.

\acknowledgments K.H. is supported in part by IARPA and 
a grant of HPC resources from the DOD HPCMP. O.T.V. thanks
IARPA for grant N66001-12-1-2023. M.A. thanks J. Linder and T. Yokoyama for
valuable discussions.

\end{document}